%% file: ours.tex
\documentclass{ar3e}
\normalsize
\newskip\zatskip \zatskip=0pt plus0pt minus0pt
\def\matth{\mathsurround=0pt}
\def\lsim{\mathrel{\mathpalette\atversim<}}
\def\gsim{\mathrel{\mathpalette\atversim>}}
\def\atversim#1#2{\lower0.7ex\vbox{\baselineskip\zatskip\lineskip\zatskip
  \lineskiplimit
0pt\ialign{$\matth#1\hfil##\hfil$\crcr#2\crcr\sim\crcr}}}
\def\inpb{\ifmmode {\rm pb}^{-1}\else ${\rm pb}^{-1}$\fi}
\def\infb{\ifmmode {\rm fb}^{-1}\else ${\rm fb}^{-1}$\fi}
\def\epem{\ifmmode e^+e^-\else $e^+e^-$\fi}
\def\ppb{\ifmmode \bar pp\else $\bar pp$\fi}
\def\pbp{\ifmmode ~^(\bar p^)p\else $~^(\bar p^)p$\fi}
\def\ie{{\it i.e.}}

\def\vereq#1#2{\lower3pt\vbox{\baselineskip1.5pt \lineskip1.5pt}}

\begin{document}
\input epsf.tex
\input psfig.sty
\def\met{\mbox{$\not\!\!{E}_{T}$}}
\jname{Submitted to Annu. Rev. Nucl. Part. Phys.}
\jyear{2002,~~~~~~~~~~~~~~~~~~~~~~~~~~~EFI-02-70,~SLAC-PUB-9194}
\vskip-0.2cm
\title{Particle Physics Probes Of Extra Spacetime Dimensions}
\markboth{JH,MS}{Particle Physics Probes Of Extra Spacetime Dimensions}
\author{JoAnne Hewett$^a$, Maria Spiropulu$^b$
\affiliation{~$a$ Stanford Linear Accelerator Center, Stanford, CA  94309\\
~$b$ Enrico Fermi Institute, Chicago, IL 60637}}
\vskip-3em
\begin{keywords}
collider, gravity, spacetime, dimensions, hierarchy, braneworld
\end{keywords}
\vskip-3em
\begin{abstract}
The possibility that spacetime is extended beyond the familiar 
$3+1$-dimensions has intrigued physicists for a century.  Indeed,
the consequences of a dimensionally richer spacetime would be profound.
Recently, new theories with higher dimensional spacetimes 
have been developed to resolve the hierarchy problem in 
particle physics.  These scenarios make distinct predictions which allow
for experiment to probe the existence of extra dimensions in new ways.
We review the conceptual framework of these scenarios, their
implications in collider and short-range gravity experiments, 
their astrophysical and cosmological effects, as well as the 
constraints placed on these models from present data.
\end{abstract}
\vskip-0.20cm
\maketitle
\section{Introduction}
\label{intro}
\input{intro}

\subsection{General}
\label{gen}
\input{gen}

\subsection{Basic Theory Concepts}
\label{primer}
\input{primer}

\section{Principal Scenarios and Collider Constraints}
\label{setmod}
\subsection{Large Extra Dimensions}
\input{add}

\subsection{Warped Extra Dimensions with Localized Gravity}
\label{rs}
\input{rs}

\subsection{TeV$^{-1}$-Sized Extra Dimensions}
\label{ignatios}
\input{tev}

\section{Tests of Short Range Gravity }
\label{table}
\input{table}

\section{Astrophysical and Cosmological Constraints}
\label{astro}
\input{astro}
\section{Summary and Extra Directions}
\label{sum}
\input{sum}

\section{Acknowledgements}
MS is supported in part by NSF
grant PHY 00-70764 and the Pritzker Foundation.
JH is supported by the Department of Energy, Contract
DE-AC03-76SF00515.
\section{Bibliography}

\input{bib}
\end{document}

%% file: intro.tex
The possible existence of additional dimensions beyond our
usual $3+1$-dimensional universe captivates
the imagination.  String theorists and their predecessors have
studied the effects of higher dimensional spacetimes at  very 
short distances
for decades.  New theories, motivated by one of the most puzzling
questions in particle physics, have proposed that the effects
of extra dimensions might be visible at larger distances, comparable
to the TeV energy scale.  This allows for present day experiments
to explore their presence.  Here, we review the principal features
of these scenarios and their experimental consequences.

%% file: gen.tex
In particle physics, 4-dimensional Minkowski spacetime is
the underlying fundamental framework under which the laws of nature
are formulated and interpreted.  Relativistic quantum fields
exist in spacetime, interactions occur at  spacetime points
and the laws governing these fields and their interactions
are constructed using weighted averages over their
spacetime histories. According to the general theory of relativity,
fluctuations of the spacetime curvature provide gravitational
dynamics. Indeed, experiments show evidence for the predictions
of general relativity and hence that spacetime is dynamical
at very long length scales.  However,
gravitational dynamics have yet to be probed at short distances, and
it is possible that they are quite different from that implied by a simple
extrapolation of the long range theory.

Early attempts to extend general relativity in order to unify
gravity and electromagnetism within a common
geometrical framework trace back to
Gunnar Nordstr\"om (1914) \cite{gn:14},
Theodor Kaluza (1921) and Oscar Klein (1926) \cite{kk:21}.  They
proposed that unification of the two forces occurred when spacetime 
was extended to a five dimensional  manifold and imposed the
condition that the fields should not depend on the
extra dimension. A difficulty with the acceptance of these ideas 
at the time was a lack of both
experimental implications and a quantum description of gravitational dynamics.

Today, one of the most striking requirements of modern string theory,
which
incorporates both gauge theories and gravitation,
is that  there must be six or seven extra spatial dimensions. Otherwise
the theory is anomalous.  Recently, concepts
developed within string theory have led to new phenomenological ideas
which relate the physics of extra dimensions to observables in a
variety
of physics experiments.

These new theories have been developed to address the hierarchy
problem, \ie, the large disparity between the electroweak scale
($\sim 10^3$ GeV) where electroweak symmetry breaking occurs 
and the traditional scale of gravity defined by the
Planck scale ($10^{19}$ GeV).
The source of physics which generates and stabilizes this sixteen order
of magnitude difference between the two scales is unknown and 
represents one of the most puzzling aspects of nature.
The novel approach
to this long-standing problem proposed in
these recent theories is that the geometry of extra spatial dimensions
may be responsible for the hierarchy: the gravitational field lines 
spread throughout the full higher dimensional space and modify the
behavior of gravity.  Indeed,
the fact that gravity has yet to be measured at energy scales much
above $10^{-3}$ eV in laboratory experiments admits for the possibility
that at higher energies gravity behaves quite differently than expected.
The first scenario of this type to be proposed \cite{ar:98}
suggested that the apparent
hierarchy between these two important scales of nature is generated
by a large volume of the extra dimensions, while in a later
theoretical framework \cite{ra:99,Randall:1999vf} 
the observed hierarchy results from a
strong curvature of the extra dimensional space.
If new dimensions are indeed relevant to the source of the hierarchy,
then they should provide detectable signatures at the electroweak
scale. These physics scenarios with
additional dimensions hence afford concrete and distinctive
phenomenological predictions for high energy colliders, as well as
producing observable consequences for astrophysics and short-range
gravity experiments.

%% file: primer.tex
Theoretical frameworks with extra dimensions have some general
features. In most  scenarios, our observed $3$-dimensional space
is  a 3-brane (sometimes called a wall), where the
terminology is derived from a generalization of a 2-dimensional
membrane.  This 3-brane
is embedded in a  higher $D$-dimensional spacetime, $D=3+\delta+1$, with
$\delta$ extra spatial dimensions which are orthogonal to our 3-brane.
The higher $D$-dimensional space is known as the ``bulk''.
The branes provide a mechanism
to hide the existence of extra dimensions in that an
observer trapped on a brane  can not directly probe
the dimensions transverse to the brane
without overcoming the brane tension.    
String theory contains branes upon which particles
can be naturally confined or localized \cite{pol}.  
In a general picture, branes carry the Standard Model gauge
charges and the ends of open strings are stuck to the branes 
and represent the 
Standard Model fields.   Fields, such as gravitons, 
which do not carry Standard Model gauge charges
correspond to closed strings and
may pop off the brane and propagate throughout the bulk.

The picture is thus one where matter and 
gauge forces
are confined to our 3-dimensional subspace, while gravity propagates
in a higher dimensional volume.  In this case,
the Standard Model fields maintain their usual behavior, however,
the
gravitational field spreads throughout the full $3+\delta$ 
spatial volume.  Conventional wisdom dictates that if the 
additional dimensions are too large, 
this would result in observable deviations from Newtonian gravity.  
The extra dimensional space must then be {\it compactified}, {\it i.e}, 
made finite.  However, in some alternative theories 
\cite{Randall:1999vf,Arkani-Hamed:1999hk},
the extra dimensions are 
infinite and the gravitational deviations are suppressed by
other means.

If the additional dimensions are small enough, the Standard Model
fields are phenomenologically allowed to propagate in the bulk.
This possibility allows for new model-building techniques to
address gauge coupling unification \cite{dudas1}, 
supersymmetry breaking \cite{Antoniadis:1990ew,Kaplan:1999ac}, the
neutrino mass spectrum \cite{Dienes:1998sb}, 
and the fermion mass hierarchy \cite{Arkani-Hamed:2000dc}. 
Indeed, the field content which is allowed to propagate in the
bulk, as well as the size and geometry of the bulk itself,
varies between different models.  

As a result of compactification,
fields propagating in the bulk expand into a series of 
states known as
a Kaluza-Klein (KK) tower, with the individual KK excitations being
labeled by mode numbers.
Similar to a particle in a box, the momentum of the bulk field is
then quantized in the compactified dimensions.
For an observer trapped on
the brane, each quanta of momentum in the compactified volume
appears as a KK excited state with mass $m^2=\vec p_\delta^{~2}$.  This
builds a KK tower of states, where each state carries identical spin
and gauge quantum numbers.  
If the additional dimensions are infinite instead of being compactified,
the $\delta$-dimensional momentum and resulting KK spectrum is
continuous.

More technically, in the case where gravity propagates in a
compactified  bulk,
one starts from a $D$-dimensional Einstein-Hilbert
action and performs a KK expansion 
about the metric field of the higher dimensional spacetime.
The graviton KK towers arise as a solution to the linearized equation of motion
of the metric field in this background \cite{tasi:00}. 
The resulting 4-dimensional fields are the 
Kaluza-Klein modes.  Counting the degrees of freedom within 
the original higher dimensional metric, the reduction of a spin-2 bulk field
results in three distinct classes of towers of KK modes:
symmetric tensor,  vector fields and scalar fields. 
The KK zero-mode 
fields are massless, while the excitation states acquire mass by
`eating' lower spin degrees of freedom.
This results in a single 5-component
tensor KK tower of massive graviton states, 
$\delta-1$ gauge KK towers of massive vector states, and $\delta
(\delta-1)/2$ scalar towers.
The zero-mode scalar states are 
radius moduli fields associated with the size of the additional
dimensions.

A generalized calculation of the
action for linearized gravity in $D$ dimensions can be used to
compute the effective 4-dimensional theory.  The spin-2 tower of
KK states couples to Standard Model fields on the brane via
the conserved symmetric stress-energy tensor.
The spin-1 KK tower does
not induce interactions on the 3-brane. The scalar KK states
couple to the Standard Model fields on the brane via the trace of
the stress-energy tensor.

The possible experimental signals for the existence of extra dimensions 
are: (i) the direct or indirect observation of a KK tower of states,
or (ii) the observation
of deviations in the inverse-square law of gravity in short-range
experiments.  The detailed properties of
the KK states are determined by the geometry of the compactified space
and their measurement would reveal the underlying geometry
of the bulk.

%% file: add.tex

The large extra dimensions scenario postulated by Arkani-Hamed, Dimopoulos
and Dvali (ADD) \cite{ar:98} makes use of the string inspired braneworld
hypothesis. In this model, the Standard Model gauge and matter fields
are confined to a 3-dimensional brane that exists within a
higher dimensional bulk. Gravity alone propagates in the
$\delta$ extra spatial dimensions which are compactified.  
Gauss' Law relates the Planck scale of the
effective 4-d low-energy theory, $M_{\rm Pl}$, to the scale
where gravity becomes strong in the $4+\delta$-dimensional spacetime,
$M_D$, through the volume of the compactified dimensions $V_\delta$ via
\begin{equation}
M^2_{\rm Pl}=V_\delta M_D^{2+\delta}\,.
\end{equation}
Taking $M_D\sim$ TeV, as assumed by ADD, eliminates the hierarchy
between $M_{\rm Pl}$ and the electroweak scale.  $M_{\rm Pl}$ is  generated 
by the large volume of the higher dimensional space and is thus no 
longer a fundamental scale.
The hierarchy problem is now translated to the possibly more tractable
question of why the compactification scale of the
extra dimensions is large.

If the compactified dimensions are
flat, of equal size, and of toroidal form,  then $V_\delta
=(2\pi R_c)^\delta$.
For $M_D \sim$ TeV, the radius $R_c$ of the extra dimensions ranges from
a fraction of a millimeter 
to  $\sim 10$ fermi for $\delta$ varying between 2 and 6.  The
compactification scale ($1/R_c$) associated with these parameters then ranges
from $\sim 10^{-4}$ eV to tens of MeV.  The case of one extra dimension
is excluded as the corresponding dimension (of size $R_c\approx 10^{11}$ m) 
would directly alter Newton's law at solar-system distances. 

Our knowledge of the electroweak and strong forces extends 
with great precision 
down to distances of
order $10^{-15}$ mm, which corresponds to $\sim(100~{\rm GeV})^{-1}$.
Thus the Standard Model fields do not
feel the effects of the large extra dimensions present in this
scenario and must be confined to the  $3$-brane.  
Therefore in this model only gravity probes the existence of the extra 
dimensions \footnote{Any Standard Model singlet field, {\it e.g.}, right
handed neutrinos, could also be in the bulk in this scenario.}.  

In particle physics, an extra dimension of size $R_c\gsim 10$ fermi 
is a large one.  If such dimensions are present and quantum gravity 
becomes strong at the
TeV scale, then observable signatures at
colliders operating at TeV energies must be induced.  
Since only gravity probes the existence of the bulk,  
is it possible that such effects  are observable in
particle collisions on the brane by means of the interactions of the bulk
graviton with the Standard Model fields? 

Technically, the 4-dimensional effective theory is
computed within linearized
quantum gravity.  The flat metric is expanded via
\begin{equation}
G_{AB}=\eta_{AB}+{h_{AB}\over M_D^{\delta/2+1}}
\end{equation}
where the upper case indices extend over the full D-dimensional
spacetime and $h_{AB}$ represents the bulk graviton fluctuation.
The interactions of the graviton are then described by the action
\begin{equation}
S_{int} = -{1\over M_D^{\delta/2+1}}\int d^4x\, d^\delta y_i\,
h_{AB}(x_\mu,y_i)T_{AB}(x_\mu,y_i)\,,
\end{equation}
with $T_{AB}$ being
the symmetric conserved stress-energy tensor.
Upon compactification, the bulk graviton decomposes into the
various spin states as described in the introduction and 
Fourier expands into Kaluza-Klein towers of spin-0, 1, and 2 
states which have equally spaced 
masses of $m_{\vec n}=\sqrt{\vec n^2/R_c^2}$,
where $\vec n=(n_1,n_2,...n_\delta)$ labels the KK excitation level.
The spin-1 states do not interact with fields on the 3-brane, the
spin-0 states couple to the trace of the stress-energy tensor and
will not be considered here. Their phenomenology is described in
\cite{grw2}.  Performing the KK expansion for the spin-2 tower,
setting $T_{AB}=\eta^\mu_A\eta^\nu_BT_{\mu\nu}\delta(y_i)$
for the Standard Model fields confined to the brane, and integrating 
the action over the extra dimensional
coordinates $y_i$ gives the interactions of the graviton KK states
with the Standard Model fields.  All the states in the 
KK tower, including the $\vec n=0$ massless state, couple in an
identical manner with universal strength of $M_{\rm Pl}^{-1}$.
The corresponding Feynman rules are catalogued in
\cite{Giudice:1999ck,Han:1999sg}.

One may wonder how interactions of this type can 
be observable at colliders since the coupling strength is so weak. 
In the ADD scenario, there are $(ER_c)^\delta$ massive Kaluza-Klein
modes that are kinematically accessible in a collider process with
energy $E$.  For $\delta=2$ and $E=1$ TeV, that totals $10^{30}$ graviton
KK states which may individually contribute to a process.  It is the
sum over the contribution from each KK state which removes 
the Planck scale suppression in a process and replaces it by powers of the 
fundamental scale $M_D\sim$ TeV.  The interactions of
the massive  Kaluza-Klein graviton modes 
can then be observed in collider experiments either
through missing energy signatures or through their virtual exchange in
Standard Model processes. At future colliders with very high energies,
it is possible that quantum gravity phenomena are accessible resulting
in explicit signals for string or brane effects; these will be discussed
briefly in Section 5.  We now discuss in detail the two classes of collider
signatures for large extra dimensions.

The first class of collider processes involves the real emission of
Kaluza-Klein graviton states in the scattering 
processes $\epem\to\gamma(Z)+G_n$, 
and $\pbp\to g+G_n$, or in $Z\to f\bar f+G_n$.  
The produced graviton  behaves as if it were a massive,
non-interacting, stable particle and thus appears 
as missing energy in the detector.  The cross
section is computed for the production of a single massive KK excitation
and then summed over the full tower of KK states.  
Since the mass splittings
between the KK states is so small, 
the sum over the states may be
replaced by an integral weighted by the  density of KK states.  The
specific process kinematics cut off this integral, rendering a
finite and model independent result.  The expected suppression from 
the $M_{\rm Pl}^{-1}$ strength of
the graviton KK couplings is exactly compensated  by a $M^2_{\rm Pl}$
enhancement in the phase space integration.  The cross
section for on-shell production of massive Kaluza Klein graviton
modes then scales as simple powers of $\sqrt s/M_D$,
\begin{equation}
\sigma _{KK} \sim {1\over M_{\rm Pl} ^2}(\sqrt sR_{c})^\delta 
\sim {1\over M_D^2}
\left({\sqrt s\over M_D}\right)^\delta \; .
\end{equation}
The exact expression may be found in 
\cite{Giudice:1999ck,Mirabelli:1999rt}.
It is important to note that due to integrating over the effective density
of states, the radiated 
graviton appears to have a continuous mass distribution;
this corresponds to the probability of emitting gravitons with different
extra dimensional momenta.  The observables for graviton
production, such as the $\gamma/Z$ angular and energy distributions in
\epem\ collisions, are then distinct from those of other  physics processes
involving fixed masses for the undetectable particles.
 
\begin{figure}[htbp]
\centerline{
\psfig{figure=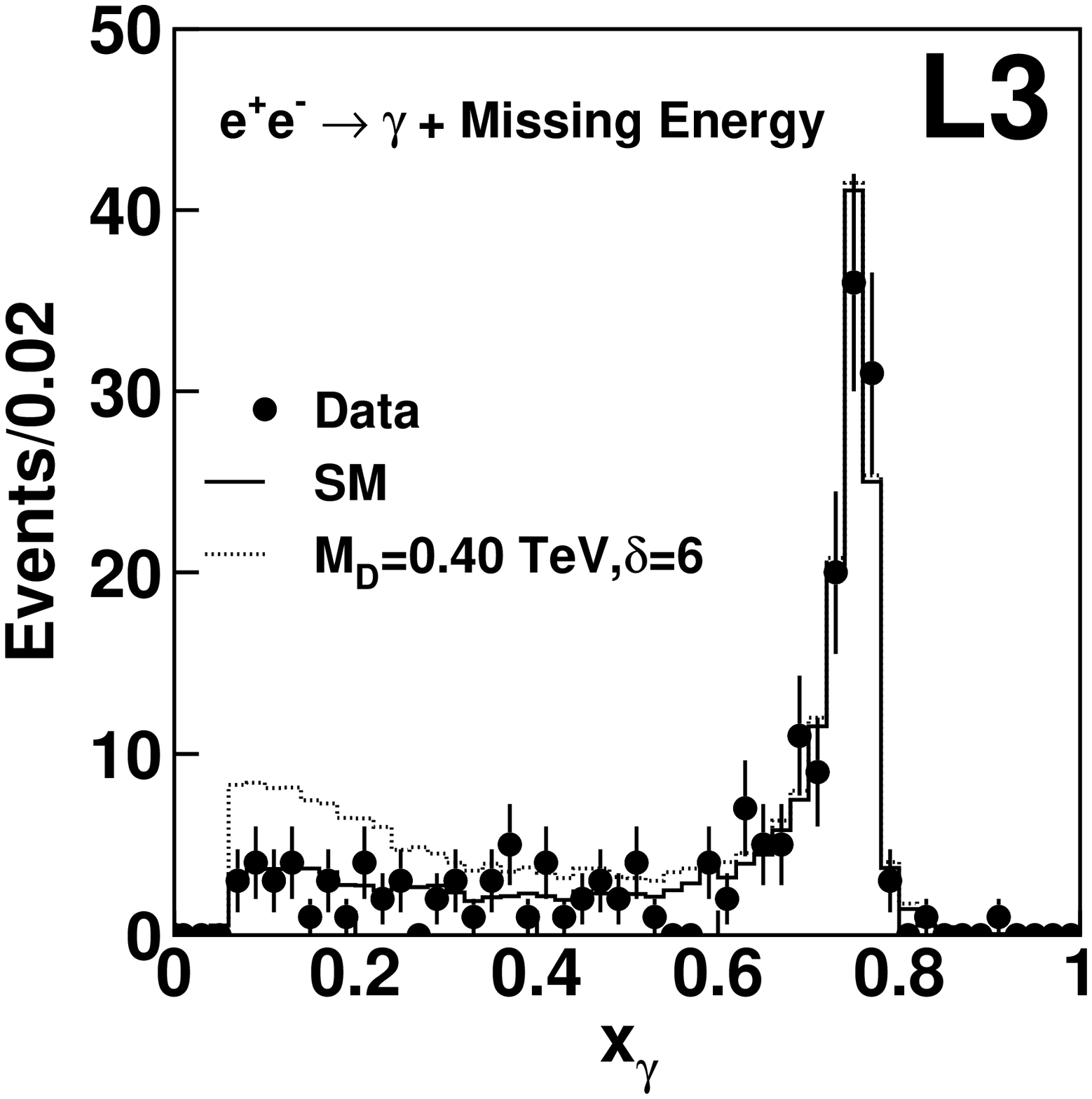,height=9cm}}
\vspace*{-.1cm}
\caption{Distribution of the ratio of the photon 
energy to the beam energy for single photon events at LEP with
$\sqrt{s}$=183 GeV.  The data and expectations from the Standard
Model for this reaction are indicated as labeled.
The effect of real 
graviton production with six extra spatial dimensions and $M_D=0.4$ TeV
is also shown. From \cite{L3:99}.}
\label{single_photon}
\end{figure}

Searches for direct KK graviton production in the reaction
$e^{+}e^{-}\to G_n+\gamma(Z)$ at LEP II, using
the characteristic final states  of  missing energy plus a
single photon or  Z boson, have excluded \cite{lepsum} fundamental scales 
up to $\sim 1.45$ TeV for two extra compactified dimensions and $\sim 0.6$ TeV
for six extra dimensions.  
These analyses use both total cross section
measurements and fits to angular distributions  to set a limit on the
graviton production rates as a function of the number of extra
dimensions.  An example of the expected photon energy distribution for
single photon production in
this scenario, compared to the data, is displayed in Fig. \ref{single_photon}
from \cite{L3:99}.

The expected discovery reach from this process
has been computed in Ref. \cite{teslatdr} at a high energy linear
\epem\ collider with 
$\sqrt s=800$ GeV, 1000 fb$^{-1}$ of integrated luminosity, and various
configurations for the beam polarization.  
These results are displayed in Table \ref{emit_tab}
and include kinematic acceptance cuts, initial state radiation, and
beamsstrahlung.  

The emission process at hadron colliders, for example, $q\bar q\to g+G_n$,
results in a monojet plus missing transverse energy  signature.
For larger numbers of extra dimenisons the density of the KK states
increases rapidly and the KK mass distribution is shifted to higher
values. This is not reflected in the missing energy distribution:
although the heavier KK gravitons are more likely to carry larger energy,
they are also more likely to be produced at threshold due to the rapidly
decreasing parton distribution functions.
These two effects compensate each other, leaving nearly identical 
missing energy distributions.  In addition, the
effective low-energy theory breaks down for some regions 
of the parameter space  as the parton-level center of mass 
energy can exceed the value of $M_D$. Experiments are then 
sensitive to the new physics appearing above 
$M_D$ that is associated with the 
low-scale quantum gravity.  

\begin{figure}[htbp]
\centerline{
\psfig{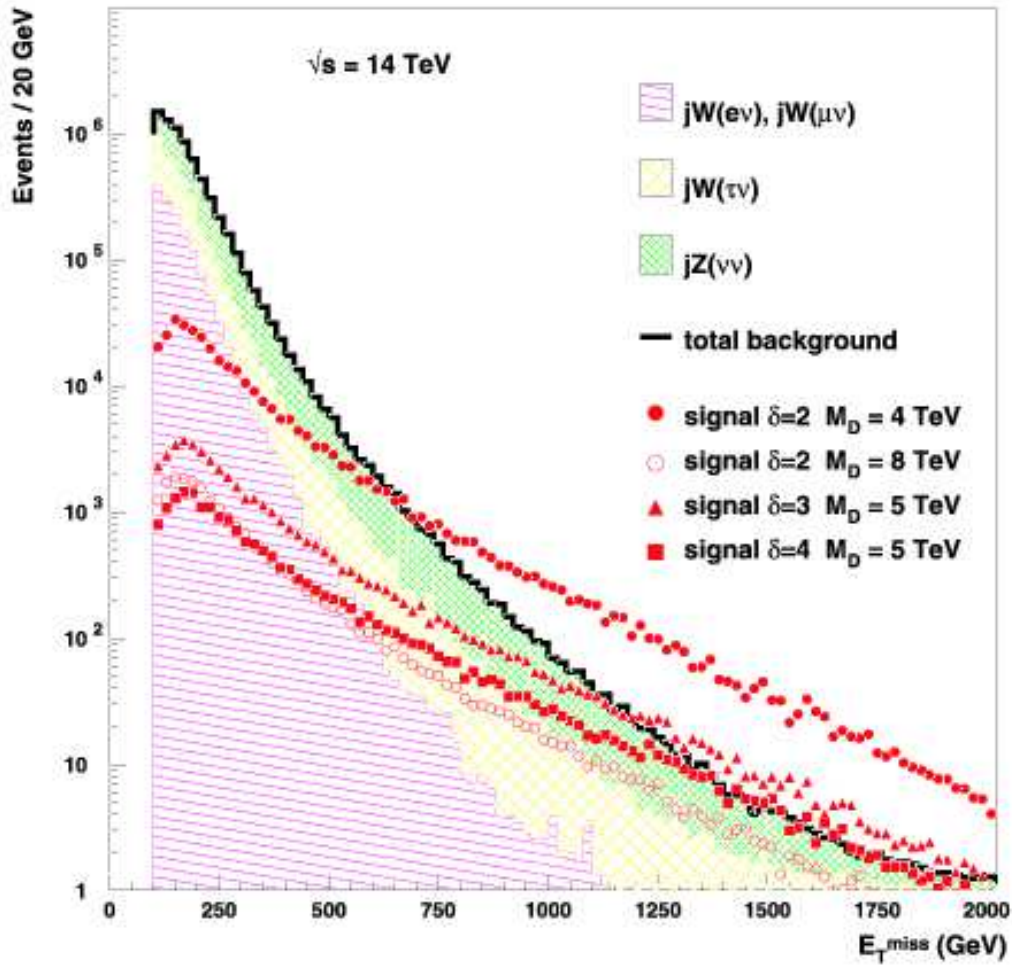}}
\vspace*{-.1cm}
\caption{Distribution of the missing transverse energy in background
events and signal events for 100 fb$^{-1}$. The contribution of the
three principal Standard Model background processes is shown as well as the
distribution of the signal for several values of $\delta$ and $M_{D}$.
From \cite{ian:01}.} 
\label{ian_mono}
\end{figure}

Searches from the Tevatron Run I are expected to yield similar results as
those from LEP II, however it is anticipated that Run II at the Tevatron 
will have a higher sensitivity \cite{Cullen:2000ef}.
An ATLAS simulation \cite{ian:01} of the missing
transverse energy in signal and background events at the LHC with 100
fb$^{-1}$ is presented in Fig. ~\ref{ian_mono} for various values of
$M_D$ and $\delta$.    This study
results in the discovery range displayed in Table
\ref{emit_tab}.   The lower end of the range corresponds to where the
ultraviolet physics sets in and the effective theory fails, while the upper
end represents the boundary where the signal is observable above background.

If an emission signal is observed, one would like to determine the values of
the fundamental parameters, $M_D$ and $\delta$.  The measurement
of the cross section at a linear collider at two different values of
$\sqrt s$ can be used to determine these parameters \cite{teslatdr} and test 
the consistency of the data with the large extra dimensions hypothesis.

Lastly, we note that the cross section for the emission process
can be reduced somewhat if the 3-brane is flexible, or soft, instead
of being rigid \cite{Murayama:2001av}.  In this case, the brane is allowed to
recoil when the KK graviton is radiated; this can be parameterized
as an exponential suppression of the cross section, with the exponential
being a function of the brane tension.  For reasonable values of 
the brane tension, this suppression is not numerically large.
The brane tension can also be determined by 
mapping out the cross section as a
function of $\sqrt s$.

\begin{table}
\centering
\begin{tabular}{|c|l|c|c|c|} \hline\hline
$e^+e^-\to\gamma+G_n$ & & 2 & 4 & 6 \\ \hline
LC & $P_{-,+}=0$ & 5.9 & 3.5 & 2.5 \\
LC & $P_{-}=0.8$ & 8.3 & 4.4 & 2.9 \\
LC  & $P_{-}=0.8$, $P_{+}=0.6$  & 10.4 & 5.1 & 3.3\\
$ pp\to g+G_n$ & & 2 & 3 & 4\\ \hline
LHC  & & $4 - 8.9$ & $4.5-6.8$ & $5.0-5.8$ \\ \hline\hline
\end{tabular}
\caption{$95\%$ CL sensitivity to the fundamental scale $M_D$ in TeV
for different values of $\delta$,
from the emission process for various polarization configurations and
different colliders as discussed in the text.  $\sqrt s = 800 $ GeV
and 1 ab$^{-1}$ has been assumed for the LC and 100 fb$^{-1}$ for the
LHC.  Note that the LHC only
probes $M_D$ within the stated range.}
\label{emit_tab}
\end{table}

The second class of collider signals for large extra dimensions is that of
virtual graviton exchange\cite{Giudice:1999ck,Hewett:1998sn}
in $2\to 2$ scattering.  This
leads to deviations in cross sections and asymmetries
in Standard Model processes, such as $e^+e^-\to f\bar f$. It may also
give rise to new production processes which are not present at tree-level in
the Standard Model, such as $gg\to\ell^+\ell^-$.  The signature is
similar to that expected in composite theories and
provides a good experimental tool
for searching for large extra dimensions for the case $\sqrt s < M_D$.

Graviton exchange is governed by the effective Lagrangian
\begin{equation}
{\cal L}=i{4\lambda\over M_H^4}T_{\mu\nu}T^{\mu\nu} + h.c.
\end{equation}
The amplitude is proportional to the sum over the
propagators for the graviton KK tower which  may be
converted to an integral over the density of KK states.  However, in this
case, there is no specific cut-off associated with the process kinematics
and the integral is divergent for $\delta>1$.  This introduces a
sensitivity to the unknown ultraviolet physics which appears at the
fundamental scale.  This integral needs to be regulated and several 
approaches have been proposed:  (i) a naive
cut-off scheme \cite{Giudice:1999ck,Hewett:1998sn}
(ii) brane fluctuations \cite{bando}, or 
(iii) the inclusion of full weakly coupled
TeV-scale string theory in the scattering process \cite{Cullen:2000ef,dudas}.
The most model independent approach which does not make any assumptions
as to the nature of the new physics appearing at the fundamental scale
 is that of the naive cut-off.  Here, 
the cut-off is set to $M_H\neq M_D$; the exact relationship between
$M_H$ and $M_D$ is not calculable without knowledge of the full theory. 
The parameter $\lambda=\pm 1$ is also usually incorporated in direct
analogy with the standard parameterization for contact interactions
\cite{Eichten:1983hw} and accounts for uncertainties associated
with the ultraviolet physics.  The substitution  
\begin{equation}
{\cal M}\sim {i^2\pi\over {M}_{\rm Pl}^2}\, \sum^\infty_{\vec n=1}
{1\over s-m_{\vec n}^2}\to {\lambda\over M_H^4}
\end{equation}
is then performed in the
matrix element for s-channel KK graviton exchange
with corresponding replacements for t- and u-channel scattering.
As above, the Planck scale suppression is removed and
superseded by powers of $M_H\sim$TeV.

\begin{figure}[htbp]
\centerline{
\psfig{figure=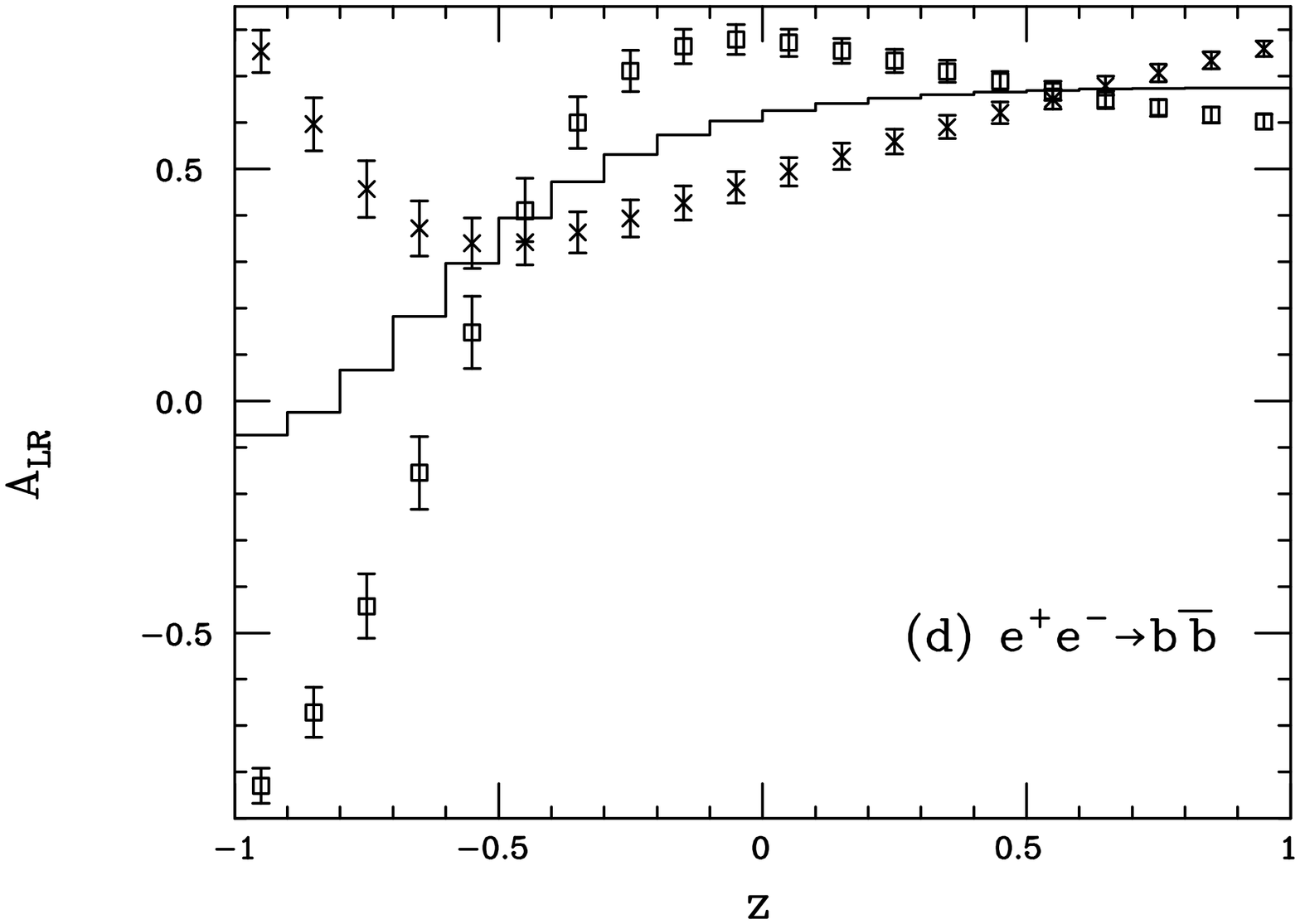,height=7.5cm,width=7.cm,angle=-90}
\hspace*{-5mm}
\psfig{figure=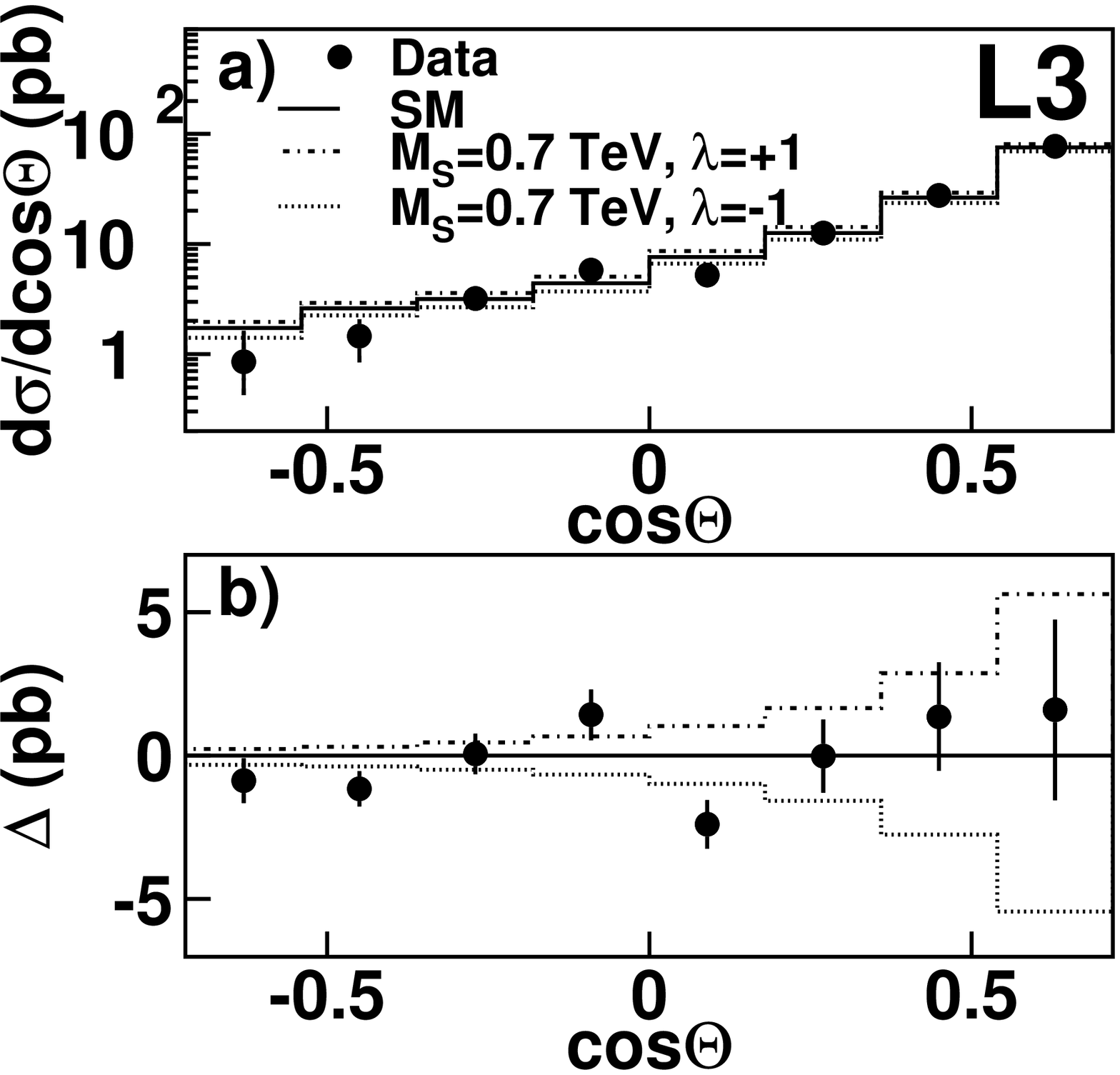,height=7.cm,width=6cm}}
\caption {(Left) Distribution of the angular dependence ($z=\cos
\theta$) of the polarized
Left-Right asymmetry in $e^+e^-\to b\bar b$ with $\sqrt s=500$ GeV, 
taking $M_D=1.5$
TeV and $\lambda=\pm 1$. The solid histogram is the Standard Model
expectation. The two sets of
data points correspond to the two choices of sign for $\lambda$, and
the error bars represent the statistics in each bin for an integrated
luminosity of 75 fb$^{-1}$.  From \cite{Hewett:1998sn}. (Right) (a) 
Differential
cross section for Bhabha scattering in the LEP data and for the Standard Model
expectations; the modification for $M_D$=0.7 TeV and
$\lambda=\pm 1$ is  as indicated.
(b) The difference of the above with the Standard
Model prediction.  From \cite{L3:99}.} 
\label{joanne_lr}
\end{figure}

The resulting angular distributions for fermion pair production
are quartic in $\cos\theta$ and thus provide a unique signal for
spin-2 exchange. 
An illustration of this is given in Fig. \ref{joanne_lr} which displays
the angular dependence of the polarized Left-Right asymmetry in 
$\epem\to b\bar b$.

The experimental analyses also make use of the cut-off approach.
Using virtual Kaluza-Klein graviton exchange in reactions with
diphoton, diboson and dilepton final states,
$e^{+}e^{-}\to G_n \to \gamma\gamma, VV, \ell\ell$,
LEP experiments \cite{lep_exch} 
exclude $M_H\lsim 0.5-1.0$ TeV independent of the number
of extra dimensions. At the Tevatron \cite{d0_ed}, the combined  Drell-Yan and
diphoton channels exclude exchange 
scales up to $\sim 1.1$ TeV.  In addition, H1 and ZEUS at 
HERA \cite{hera_ed} have both placed the bound $M_H\gsim 800$ GeV.

The potential search reach for virtual KK graviton exchange in processes 
at future accelerators are listed in Table \ref{tab_exch}.  
These sensitivities 
are estimated for the LHC \cite{lhced}, a high energy \epem\ linear 
collider \cite{Hewett:1998sn},
as well as for a $\gamma\gamma$ collider \cite{tgred},
where the initial photon beams originate from Compton laser back-scattering.
Note that the $\gamma\gamma\to WW$ process has the highest sensitivity 
to graviton exchange.

\begin{table}
\centering
\begin{tabular}{|c|c|c|c|} \hline\hline
 & & $\sqrt s$ (TeV)  & $M_H$ (TeV) \\ \hline
LC & $e^+e^-\to f \bar f$ & 0.5 & 4.1 \\
   & $e^+e^-\to f\bar f$    & 1.0 & 7.2 \\
   & $\gamma\gamma\to\gamma\gamma$ & 1.0 & 3.5\\
   & $\gamma\gamma\to WW$ & 1.0 & 13.0 \\
   & $e\gamma\to e\gamma$ & 1.0 & 8.0 \\
LHC & $ pp\to \ell^+\ell^-$ & 14.0 & 7.5 \\ 
    & $pp\to\gamma\gamma$ & 14.0 & 7.1 \\ \hline\hline
\end{tabular}
\caption{The estimated $95\%$ CL search reach for $M_H$ from various 
processes at future accelerators.}
\label{tab_exch}
\end{table}

In summary, present facilities have searched for large
extra dimensions and excluded their existence for fundamental scales
up to $\sim$ TeV.  The reach of future facilities will extend this reach
to a sensitivity of $\sim 10$ TeV.  If this scenario is indeed relevant to
the hierarchy, then it should be discovered in the next round of experiments.
In addition, future experiments will have the capability to determine
the geometry of the higher dimensional space, such as the size and number
of extra dimensions, as well as the degree of  the brane tension.

%% file: rs.tex

In this scenario,  the hierarchy between the Planck and electroweak
scales is generated by a large curvature of the extra 
dimensions \cite{ra:99,Randall:1999vf}.
The simplest such framework is comprised of just one additional
spatial dimension of finite size, in which gravity  propagates.  
The geometry is that of a  5-dimensional Anti-de-Sitter
space (AdS$_5$), which is a space of constant negative curvature.
The extent of the 5$^{th}$ dimension is $y=\pi R_c$.  Every slice
of the 5$^{th}$ dimension corresponds to a
4-d Minkowski metric.  Two 3-branes, with equal and 
opposite tension, sit at the boundaries
of this slice of AdS$_5$ space.  The Standard Model fields are 
constrained to the 3-brane located at the boundary
$y=\pi R_c$, known as the TeV-brane, while gravity is localized about
the opposite brane at the other boundary $y=0$.  This is referred to as
the Planck brane.  

The metric for this scenario preserves 4-d Poincare invariance and is
\begin{equation}
ds^2=e^{-2ky}\eta_{\mu\nu}dx^\mu dx^\nu-dy^2\,,
\end{equation}
where the exponential function of the 5$^{th}$ dimensional coordinate
multiplying the usual 4-d 
Minkowski term indicates a non-factorizable geometry.  This exponential
is known as a warp factor.
Here, the parameter $k$ governs the degree of curvature of the 
AdS$_5$ space; it is assumed to be of order the Planck scale.  Consistency
of the low-energy theory sets $k/\overline M_{\rm Pl}\lsim 0.1$, with
$\overline M_{\rm Pl} = M_{\rm Pl}/\sqrt{8\pi}=
 2.4\times 10^{18}$ being the reduced 4-d Planck scale.
The relation 
\begin{equation}
\overline M^2_{\rm Pl}={\frac{\overline M_5^3}{k}}\,
\end{equation}
is derived from the 5-dimensional action and indicates that the
(reduced) 5-dimensional fundamental scale $\overline M_5$ is of order
$\overline M_{\rm Pl}$.  Since $k\sim \overline M_5\sim \overline
M_{\rm Pl}$, there are no additional hierarchies present in this model.

The scale of
physical phenomena as realized by a 4-dimensional flat metric transverse to
the 5$^{th}$ dimension is specified by the exponential
warp factor.   The scale $\Lambda_\pi\equiv\overline{M}_{\rm Pl}
e^{-kR_c\pi}$ then describes the scale of all physical processes
on the TeV-brane.   With  
the gravitational wavefunction being localized on the 
Planck brane, $\Lambda_\pi$ takes on the value $\sim 1$ TeV
providing $kR_c\simeq 11-12$.  
It has been demonstrated \cite{gw} that this 
value of $kR_c$ can be stabilized within this configuration  
without the fine tuning of 
parameters.  The hierarchy is thus naturally established 
by the warp factor.  Note that since $kR_c\simeq 10$ and it is assumed
that $k\sim 10^{18}$ GeV, this is not a model with a large extra
dimension.

Two parameters govern the 4-d effective theory of this 
scenario \cite{dhr1}: $\Lambda_\pi$ and the ratio
$k/\overline M_{Pl}$.  Note that the approximate values of these parameters
are known due to the relation of this model to the
hierarchy problem.  As in the case of large extra dimensions, the
Feynman rules are obtained by a linear expansion of the flat
metric,  
\begin{equation}
G_{\alpha\beta}=e^{-2ky}(\eta_{\alpha\beta}
+2h_{\alpha\beta}/M_5^{3/2})\,, 
\end{equation}
which for 
this scenario includes the warp factor multiplying the linear
expansion.  
After compactification, the resulting KK tower states are the
coefficients of a Bessel expansion with the Bessel functions replacing
the Fourier series of a flat geometry
due to the strongly curved space and the presence of
the warp factor.  
Here, the masses of the KK states are $m_n=x_nk
e^{-kR_c\pi}=x_n\Lambda_\pi k/\overline M_{\rm Pl}$ with the $x_n$ being
the roots of the first-order Bessel function, {\it i.e.}, $J_1(x_n)
=0$.  The first excitation is then naturally of order a 
TeV and the KK states are not evenly spaced.  The
interactions of the graviton KK tower with the Standard Model fields on
the TeV-brane are given by
\begin{equation}
{\cal L}={-1\over\overline M_{Pl}}T^{\mu\nu}(x)h^{(0)}_{\mu\nu}(x)
-{1\over\Lambda_\pi}T^{\mu\nu}(x)\sum^\infty_{n=1}h^{(n)}_{\mu\nu}(x)
\,.
\end{equation}
Note that the zero-mode decouples and that the couplings of the excitation
states are inverse TeV strength.  This results in a
strikingly different phenomenology than in the case of large 
extra dimensions.  

\begin{figure}[t]
\centerline{
\psfig{figure=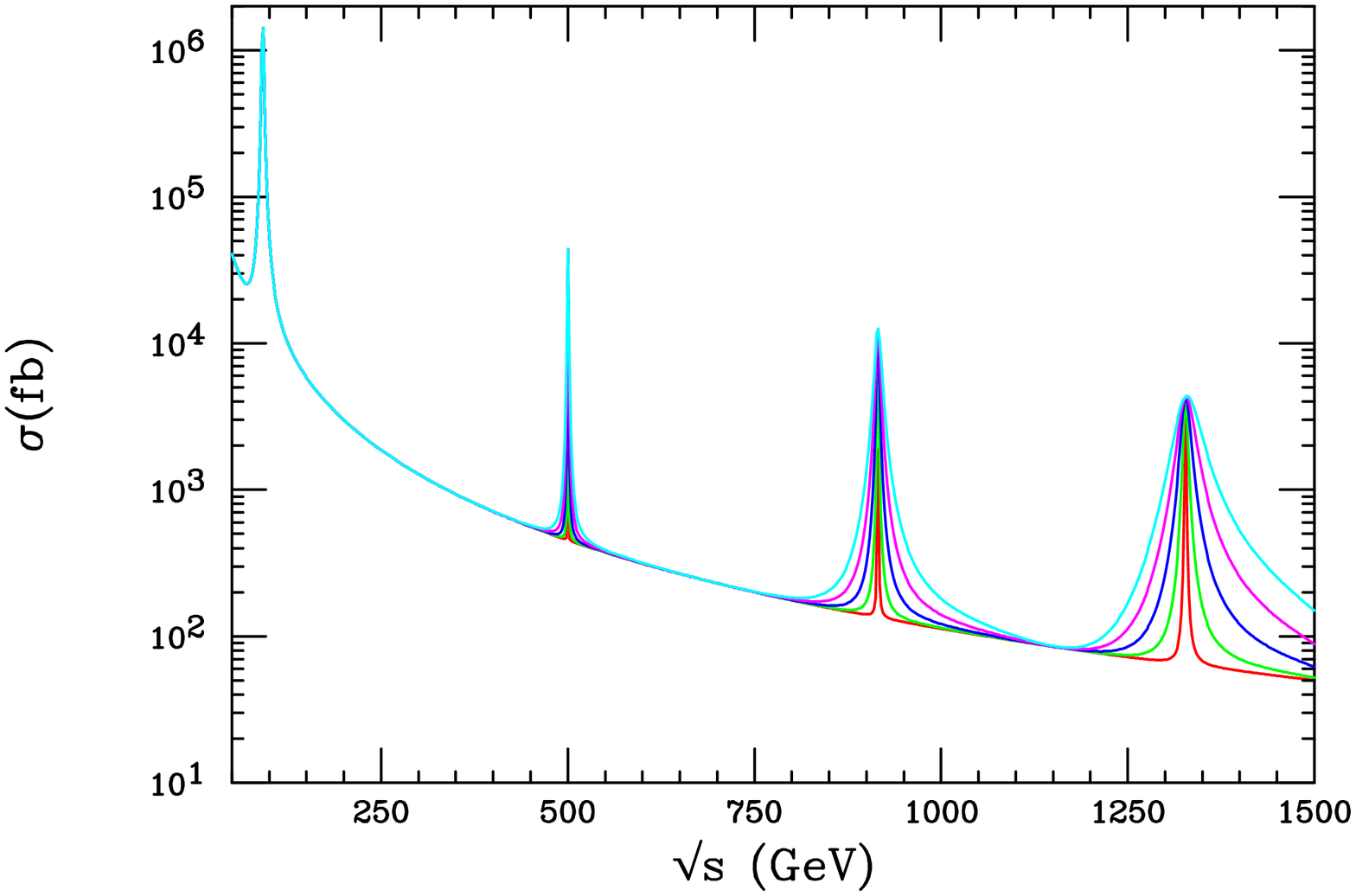,height=9cm,width=12cm,angle=90}}
\vspace*{0.1cm}
\caption{The cross section for $e^+e^-\to\mu^+\mu^-$ including the
exchange of a KK tower of gravitons in the Randall-Sundrum model 
with $m_1=500$ GeV.  The curves 
correspond to $k/\overline M_{\rm Pl}=$ in the range $0.01 - 0.05$.}
\label{kkspect}
\end{figure}

\begin{figure}[htbp]
\centerline{
\psfig{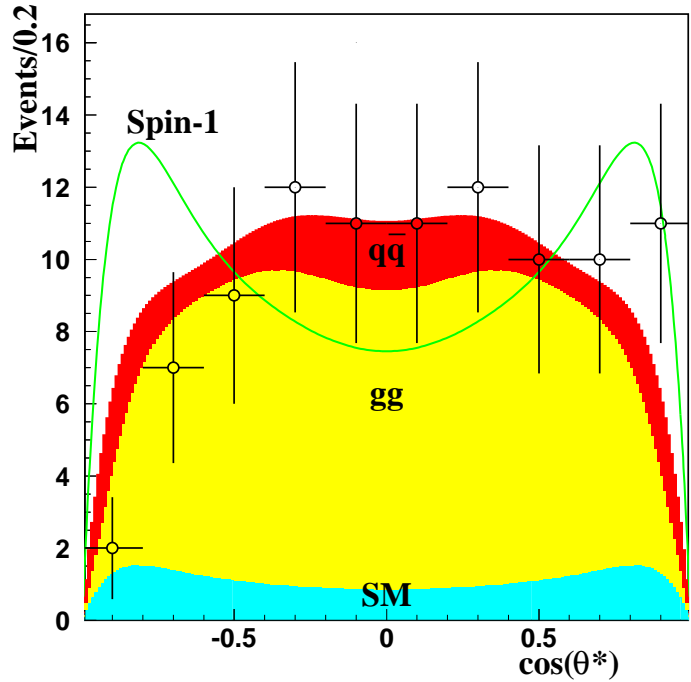}}
\vspace*{-.1cm}
\caption{The angular distribution of ``data'' 
at the LHC from Drell-Yan production
of the first graviton KK excitation with $m_1=1.5$ TeV
and 100 \infb\ of integrated luminosity.  The stacked histograms
represent the Standard Model contributions, and $gg$ and
$q\bar q$ initiated graviton production as labeled.  The curve shows the
expected distribution from a spin-1 resonance.  From \cite{britts}.}
\label{atlas_spin2}
\end{figure}

In this scenario,
the principal collider signature is the direct resonant production 
of the spin-2 states in the graviton KK tower.
To exhibit how this may appear at a  
collider, Figure  \ref{kkspect} displays the cross section for $e^+e^-\to
\mu^+\mu^-$ as a function of $\sqrt s$, assuming $m_1=500$ GeV and varying
$k/\overline M_{\rm Pl}$ in the range $0.01-0.05$.  
The height of the third resonance is greatly
reduced as the higher KK excitations prefer to decay to the lighter
graviton states, once it is kinematically allowed \cite{dr}.  In this case,
high energy colliders may become graviton factories!
If the first graviton KK state is observed, then the parameters of this 
model can be uniquely determined by measurement of the location and width
of the resonance.  
In addition, the spin-2 nature of the graviton resonance can be
determined from the shape of the angular distribution of the decay products.
This is demonstrated in Figure~\ref{atlas_spin2}, which displays the
angular distribution of the final state leptons in Drell-Yan production,
$pp\to\ell^+\ell^-$, at the LHC \cite{britts}.  

\begin{figure}[t]
\centerline{
\psfig{figure=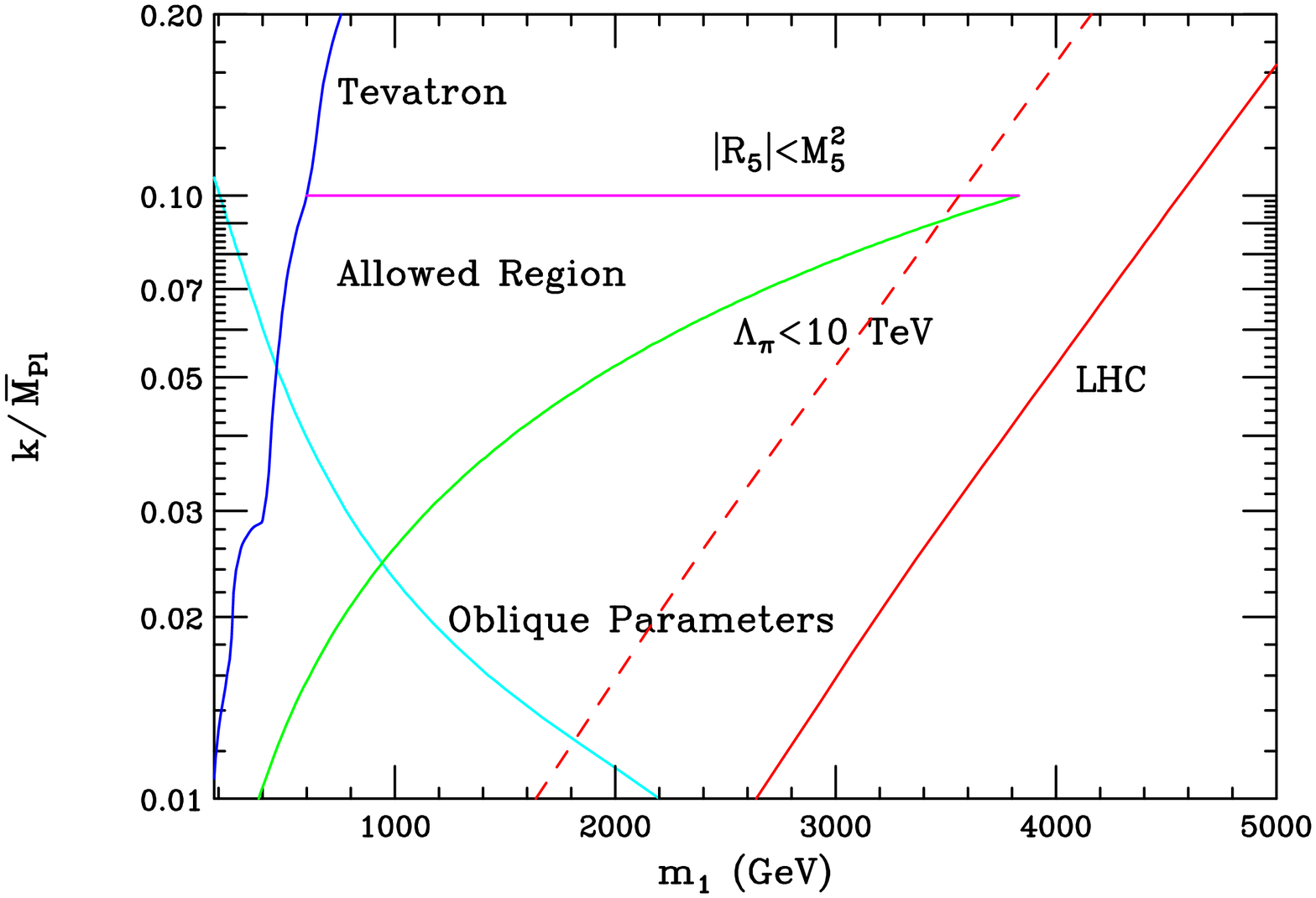,height=9cm,width=12cm,angle=90}}
\vspace*{0.1cm}
\caption{Summary of experimental and theoretical constraints on the
Randall-Sundrum model in the two-parameter plane  $k/\overline M_{\rm Pl} -
m_1$, for the case where the Standard Model fields
are constrained to the TeV-brane.  The allowed region lies in the center
as indicated.  The LHC sensitivity to graviton resonances in the
Drell-Yan channel is represented by the diagonal dashed and solid curves,
corresponding to 10 and 100 \infb\ of integrated luminosity, 
respectively.  From \cite{dhr3}.}
\label{rslhc}
\end{figure}

Searches for the first graviton KK resonance in Drell-Yan and dijet data
from Run I at the Tevatron restrict \cite{dhr1}
the parameter space of this model, as shown in Figure  \ref{rslhc}.
These data exclude larger values of $k/\overline M_{\rm Pl}$ for values
of $m_1$ which are in kinematic reach of the accelerator.

Gravitons may also contribute to precision electroweak observables.
A precise description of such contributions requires a
complete understanding of the full underlying theory due to the
non-renormalizability of gravity.  However, naive estimates
of the size of such effects can be obtained in an effective field
theory by employing a cut-off to regulate the theory \cite{cprs}.
The resulting cut-off dependent constraints indicate \cite{dhr3}
that smaller values of $k/\overline M_{\rm Pl}$ are inconsistent with
precision electroweak data, as shown in Figure \ref{rslhc}.

These two constraints from present data, taken together with the 
theoretical assumptions that
(i) $\Lambda_\pi\lsim 10 $ TeV, {\it i.e.}, the scale of physics on the 
TeV-brane is not far above the electroweak scale so that an additional
hierarchy is not generated, and (ii) $k/\overline M_{\rm Pl}
\lsim 0.1$ from bounds on the curvature of the AdS$_5$,
result in a closed allowed region in the two parameter space.
This is displayed in Figure  \ref{rslhc}, which also shows the expected
search reach for resonant graviton KK production in the Drell-Yan
channel at the LHC.  We see that the full allowed parameter space
can be completely explored at the LHC, given our theoretical
prejudices, and  hence the LHC will either discover or exclude this
model. 

If the above theoretical assumptions are evaded,
the KK gravitons may be too massive to be produced directly.
However, their contributions to fermion
pair production may still be felt via virtual exchange.  In this
case, the uncertainties associated with the
introduction of a cut-off are avoided, since there is only one additional 
dimension and the KK states may be neatly summed.  
The resulting sensitivities \cite{dhr1} 
to $\Lambda_\pi$ at current and future colliders are
listed in Table \ref{rscont}.

\begin{table}
\centering
\begin{tabular}{|c|c|c|c|} \hline\hline
 & \multicolumn{3}{c|}{$k/\overline M_{Pl}$ } \\ \hline
 & 0.01 & 0.1 & 1.0\\ \hline
LEP II & 4.0 & 1.5 & 0.4 \\
LC $\sqrt s=0.5$ TeV & 20.0 & 5.0 & 1.5 \\
LC $\sqrt s=1.0$ TeV & 40.0& 10.0 & 3.0 \\
Tevatron Run II & 5.0 & 1.5 & 0.5 \\
LHC & 20.0 & 7.0 & 3.0 \\ \hline\hline
\end{tabular}
\caption{$95\%$ CL search reach for $\Lambda_\pi$ in TeV in the contact 
interaction
regime taking 500, 2.5, 2, and 100 fb$^{-1}$ of integrated luminosity at
the LC, LEP II, Tevatron, and LHC, respectively.  From \cite{dhr1}.}
\label{rscont}
\end{table}

The Goldberger-Wise mechanism \cite{gw} for stabilizing the separation
of the two 3-branes in this configuration with $kR_c\sim 10$ 
leads to the
existence of a new, relatively light scalar field.  This field is the 
radion and it is related to the radial fluctuations of the extra
dimension, and to the scalar remnant of the bulk graviton KK
decomposition.
The radion couples to the Standard Model fields via
the trace of the stress energy tensor with strength 
$\sim T_\mu^\mu/\Lambda_\pi$.
These interactions are similar to those of the Higgs boson,
and it is allowed to mix with
the Higgs, which alters the couplings of both fields.  The phenomenology
of this field is detailed in \cite{grw2,hr}.

In a variant of this model, the Standard Model fields  may 
propagate in the bulk.  This is desirable for numerous model
building reasons as mentioned in the introduction.
As a first step, one can 
study the effect of placing the Standard Model gauge
fields in the bulk and keeping the fermions 
on the TeV-brane.  In this case, one finds \cite{dhr2} that the
fermions on the brane couple to the KK gauge fields 
$\sim 9$ times more strongly than they
couple to the Standard Model gauge fields.  This results in
strong bounds on gauge KK states from their contributions to
electroweak precision data.  A global fit to the electroweak
data set  yields the constraint $m_1
 \gsim 25$ TeV on the first gauge KK mass, implying $\Lambda_\pi\gsim 100$
TeV.

This bound can be relaxed if the fermions also reside in the 
bulk \cite{dhr3,bulkph}.
In this case, a third parameter is introduced,
corresponding to the bulk fermion mass which is given by $m_5=\nu k$
with $\nu$ being of order one.  The parameter $\nu$ 
controls the shape of the fermion zero mode wavefunction.
The resulting phenomenology is 
markedly different, and is highly dependent on the parameter $\nu$.
In particular, large mixing is induced between the zero-mode 
top-quark and the states in its KK tower.  This results \cite{hpr} in
substantial shifts to the $\rho$-parameter and forces the third
generation of fermions to be confined to the TeV-brane with only
the first two generations of fermions being allowed to reside in the bulk.

An alternate scenario is possible \cite{Randall:1999vf} 
when the second brane is taken 
off to infinity, \ie, $R_c\to\infty$, and the Standard Model fields are
confined to the brane at $y=0$ where gravity is localized.
In this case, the graviton KK modes become continuous, \ie, the gap
between KK states disappears, and their couplings to the Standard Model
fields are much weaker than $M_{\rm Pl}$.  This configuration
no longer allows for a reformulation of the hierarchy problem,
but can potentially be observable \cite{chu:00} in sub-mm 
gravitational force experiments.  

Another consistent scenario of this type involves two branes, both
with positive tension,
separated in a five-dimensional Anti-de Sitter geometry of infinite
extent \cite{lyk_randall:99}. The graviton
is localized on one of the branes, while a gapless continuum of
additional gravity modes probe the infinite fifth dimension. The
phenomenological effects of this framework are similar to the process
of real graviton emission
in the ADD scheme with six large toroidal dimensions.

%% file: tev.tex
The possibility of TeV$^{-1}$-sized extra dimensions naturally arises in 
braneworld theories \cite{Antoniadis:1990ew,more}.  
By themselves, they do not allow for a 
reformulation of the hierarchy problem, but they may be incorporated into 
a larger structure in which this problem is solved.  In these scenarios, the 
Standard Model fields are phenomenologically allowed to propagate in the 
bulk.  This presents a wide variety of choices for model building:  
(i) all, or only some, of the Standard Model gauge fields exist in the bulk;
(ii) the Higgs field may lie on the brane or in the bulk; (iii) the Standard
Model fermions may be confined to the brane or to specific locales in the
extra dimension.  The phenomenological 
consequences of this scenario strongly depend on the location of the
fermion fields.  Unless otherwise noted, our discussion
assumes that all of the Standard Model gauge fields propagate in the bulk.
 
The masses of the excitation states in the gauge boson KK towers
depend on where the Higgs boson is located. 
If the Higgs field propagates in the bulk, the zero-mode state of
the Higgs KK tower receives a  vacuum expectation value
(vev) which is responsible for the spontaneous breaking of
the electroweak gauge symmetry.  In this case, the resulting mass 
matrix for the states in the gauge boson KK towers is diagonal and
the excitation masses 
are shifted by the mass of the gauge zero-mode, which corresponds
to the Standard Model gauge field, giving
\begin{equation} 
m_{\vec n} = (m_0^2 + \vec n\cdot\vec n/R_c^2)^{1/2}\,.
\end{equation}
However, if the Higgs is confined to the brane, its vev induces mixing,
amongst the gauge KK states of order $(m_0R_c)^2$.  The KK mass
matrix must then be diagonalized in order to determine the excitation 
masses.  For the case of 1 extra TeV$^{-1}$-sized  dimension, the
coupling strength of the gauge KK states to the Standard Model fermions
on the brane is $\sqrt 2g$, where $g$ is the corresponding Standard 
Model gauge coupling.

We first discuss the case where the Standard Model fermions are
rigidly fixed to the brane and do not feel the effects of
the additional dimensions.  
For models in this class, precision electroweak data place strong
constraints \cite{tgrjim} on the mass of the first gauge KK excitation.
Contributions to electroweak observables 
arise from the virtual exchange of gauge KK states and a summation over 
the contributions from the entire KK tower must be performed.
For $D>5$, this sum is divergent.  In the full higher dimensional 
theory, some new, as of yet
unknown, physics would regularize this sum and render it finite.  An
example of this is given by the possibility that the brane is flexible
or non-rigid \cite{bando}, which has the effect of exponentially damping
the sum over KK states.  Due to our present lack of knowledge of the 
full underlying theory, the KK sum is usually
terminated by an explicit cut-off, which provides a naive estimate
of the magnitude of the effects.  

Since the $D=5$ theory is finite, it
is  the scenario that is most often discussed 
and is sometimes referred to as the 
5-dimensional Standard Model (5DSM).  In this case, a global fit
to the precision electroweak data including the contributions from
KK gauge interactions
yields \cite{tgrjim} $m_1\sim R_c^{-1}\gsim 4$ TeV.  
In addition, the KK contributions to the
precision observables allow for the mass of the Higgs boson to be 
somewhat heavier than the value obtained in the Standard Model 
global fit.
Given the constraint on $R_c$ from the
precision data set, the gauge KK contributions to the anomalous 
magnetic moment of the muon are small \cite{gm2}.

Such a large mass for the first gauge KK state is beyond the direct 
reach at present accelerators, as well as a future \epem\ linear collider.
However, they can be produced as resonances at the LHC in the 
Drell-Yan channel provided $m_1 \lsim 6$ TeV.  Lepton colliders
can indirectly observe the existence of heavy gauge KK states
in the contact interaction limit via their $s$-channel exchanges. 
In this case the contribution of
the entire KK tower must be summed, and suffers the same problems
with divergences discussed above.
The resulting sensitivities to the 
gauge KK tower in the 5DSM from direct and indirect searches at 
various facilities is displayed in Table \ref{tab_tev}.

\begin{table}
\centering
\begin{tabular}{|c|c|} \hline\hline
 & $m_1$ Reach (TeV) \\ \hline
Tevatron Run II 2 fb$^{-1}$ & 1.1 \\
LHC 100 fb$^{-1}$ & 6.3 \\
LEP II & 3.1 \\
LC $\sqrt s=0.5$ TeV 500 fb$^{-1}$ & 13.0 \\
LC $\sqrt s=1.0$ TeV 500 fb$^{-1}$ & 23.0 \\
LC $\sqrt s=1.5$ TeV 500 fb$^{-1}$ & 31.0 \\ \hline\hline
\end{tabular}
\caption{$95\%$ CL search reach for the mass $m_1$ of the first KK gauge
boson excitation \cite{tgrjim}.}
\label{tab_tev}
\end{table}

We now discuss the scenario where
the Standard Model fermions are localized at
specific points in the extra TeV$^{-1}$-sized dimensions.
In this case, the fermions have narrow gaussian-like
wave functions in the extra dimensions with the width
of their wave function being much smaller than $R_c^{-1}$.
The placement of the different fermions at distinct locations
in the additional dimensions, along with the narrowness of
their wavefunctions, can then naturally suppress \cite{Arkani-Hamed:2000dc} 
operators mediating dangerous processes such as proton decay.
The exchange of gauge KK states in $2\to 2$ scattering 
processes involving initial and final state fermions is
sensitive to the placement of the fermions and can be used to
perform a cartography of the localized fermions \cite{tgr_cart},
\ie, measure the wavefunctions and locations of the fermions.
At very large energies, it is possible that the cross section for
such scattering 
will tend rapidly to zero since the  fermions' wavefunctions will
not overlap and hence they may completely
miss each other in the extra dimensions \cite{Arkani-Hamed:1999za}.

Lastly, we discuss the case of universal extra dimensions \cite{bogdan},
where all Standard Model fields propagate in the bulk, and  
branes need not be present.  Translational
invariance in the higher dimensional space is thus preserved.
This results in the tree-level conservation of the $\delta$-dimensional 
momentum of the bulk fields, which implies that
KK parity, $(-1)^n$, is conserved to all orders.  
The phenomenology of this scenario is 
quite different from the cases discussed above.  Since KK parity is 
conserved, KK excitations can no longer be produced as s-channel
resonances; they can now only be produced in pairs.  This results
in a drastic reduction of the collider sensitivity to such states, with 
searches at the Tevatron yielding the bounds \cite{bogdan,tgruniv}
$m_1\gsim 400$ GeV for two universal extra dimensions.  
The constraints from electroweak precision
data are also lowered and yield similar bounds.
Since the KK states are allowed to be relatively light,
they can produce observable effects \cite{tgruniv,loops} in 
loop-mediated processes,
such as $b\to s\gamma$, $g-2$ of the muon, and rare Higgs decays.

%% file: table.tex

Until very recently, the inverse square force law of Newtonian 
gravity had been precisely tested only down to distances 
of order a centimeter 
\cite{PDG:98,kr:99,ho:85,mi:88,lo:99}.  Such tests are 
performed by
short range gravity experiments that probe new interactions
by searching for deviations from Newtonian gravity at small
distances.  There are several parameterizations which describe
these potential deviations \cite{ck:87}; the one  most widely
used by experiments is that where the classical gravitational
potential is expanded to include a Yukawa interaction:
\begin{equation}
V(r)=-\frac{1}{M_{\rm Pl}^{2}}\frac {m_1 m_2}{r}(1 + \alpha
e^{-r/\lambda})~.
\label{eq:potential}
\end{equation}
Here, $r$ is the distance between  two masses $m_1$ and $m_2$ and is
fixed by the experimental apparatus, $\alpha$ is a dimensionless
parameter relating the strength of the additional Yukawa interaction
to that of gravity, and $\lambda$ is the range of the new interaction.
The best experimental sensitivity is achieved for the case 
$\lambda \approx r$,
with the sensitivity decreasing rapidly for smaller distances.
The experimental results are presented in the $\alpha-\lambda$ plane 
in the form of convex curves that are centered around the distance 
at which a particular experiment operates.
These short range tests are performed by Van-der-Waals 
(probing $1/r^3$ deviations) and Casimir (probing $1/r^4$ terms)
force experiments, as well as Cavendish-type detectors which directly
measure the gravitational force.

Deviations from the inverse-square gravitational force
law can occur in the scenario with large extra dimensions
or in the alternate version of the Randall-Sundrum model where
the second brane is taken to infinity.  Focusing on the first case,
the two-body potential given by Gauss' Law in the presence of 
additional dimensions (for distances $r< R_c$)  is
expressed as \cite{ar:98}:
\begin{equation}
V(r)=-\frac{1}{8\pi  M_{D}^{2+\delta}} 
\frac{m_1 m_2}{r^{\delta+1}}
\label{eq:npotential}
\end{equation}
in the conventions of \cite{Giudice:1999ck}.
When the two masses are separated by a distance $r>R_c$ and the
dimensions are assumed to be compactified  on a torus of radius $R_c$
the potential becomes:
\begin{equation}
V(r)=-\frac{1}{8\pi M_{D}^{2+\delta}} 
\frac{m_1 m_2}{R_c^{\delta}} \frac{1}{r}\,,
\label{eq:nRpotential}
\end{equation}
{\it i.e.}, the usual $1/r$ Newtonian potential is recovered using
Gauss' Law.
The parameters in the general form of the two-body
potential in Eq.  (\ref{eq:potential}), \ie,
$\alpha$ and $\lambda$, depend on the
number of extra dimensions and the type of
compactification \cite{Kehagias:1999my};
for the simple case of compactifying
on a torus, the range $\lambda$ of the new interaction
is the compactification radius $R_c$, and $\alpha=2\delta$. 
It should be noted that the dependence of Eq. (\ref{eq:nRpotential})
on $M_D$ is related to the compactification scheme through
the precise form of the volume factor.

\begin{figure}
\epsfysize=4.162in
\epsfbox{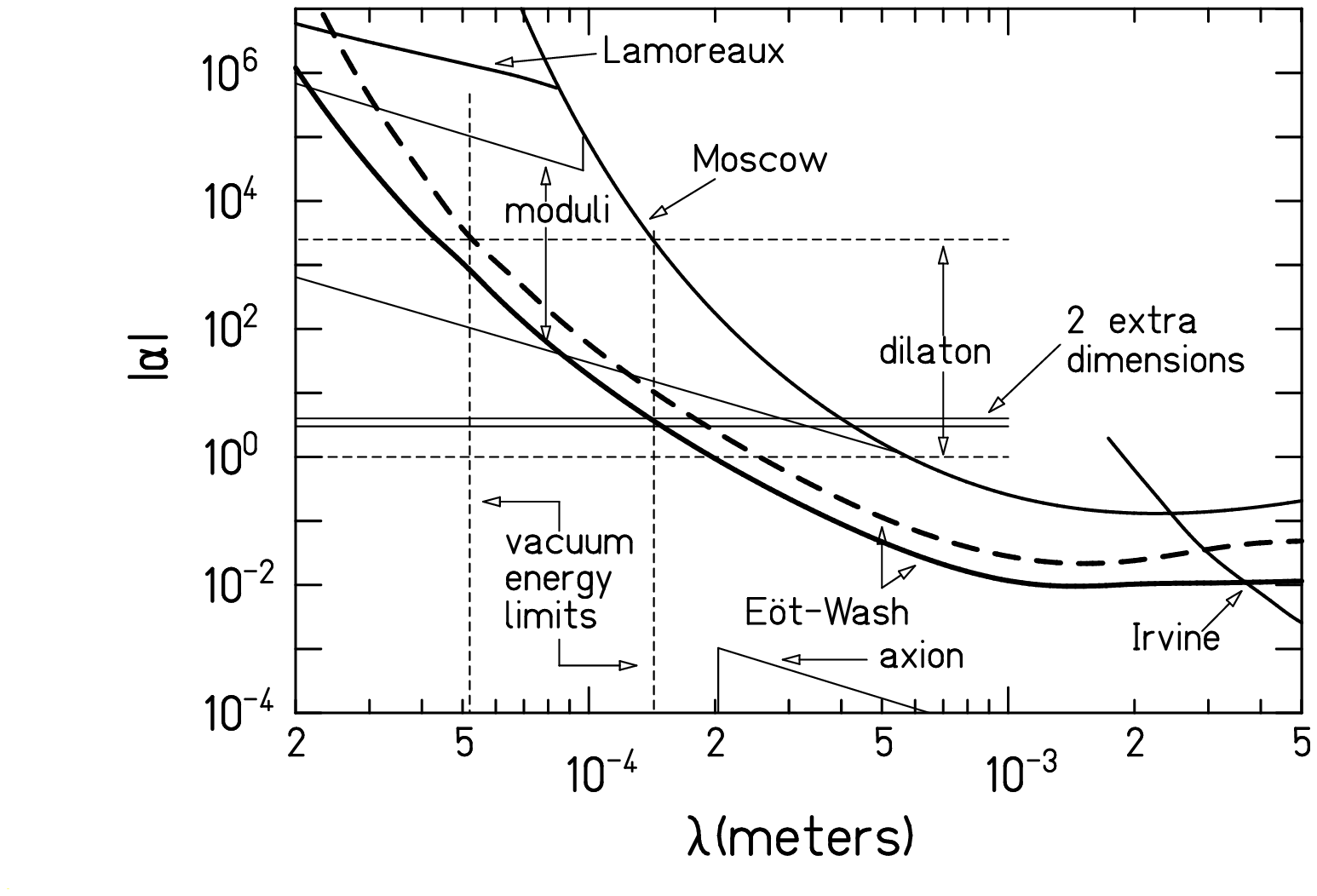}
\caption{95\% confidence level upper limits on the strength $\alpha$
(relative to Newtonian gravity) as a function of the range $\lambda$ 
of additional
Yukawa interactions.  The region excluded by previous experiments
\protect\cite{ho:85,mi:88,la:97} lies above the curves labeled Irvine,
Moscow and Lamoreaux, respectively. The most recent results
\cite{ho:00} correspond to the curves labeled E\"ot-Wash.
The heavy dashed line is the result from Ref. \cite{ho:00}, the heavy solid
line shows the most
recent analysis that uses the total data sample.}
\label{fig:adelberger}
\end{figure}

The most recent Cavendish-type experiment \cite{ho:00} 
used a torsion pendulum and a rotating attractor.  It
excludes scenarios with $\alpha \ge 3$ for $\lambda \ge 
140 ~\mu$m at 95\% confidence level.
These results are displayed as the curve labeled E\"ot-Wash in
Figure \ref{fig:adelberger}.
Interpreted within the framework of two large additional
spatial dimensions, the results
imply that $R_c < 190 ~\mu$m.  The relation of this bound
to the fundamental scale $M_D$ depends on the compactification
scheme.  For $\delta>2$, $R_c$ is too small for the effects of
extra dimensions to be probed in mechanical experiments.

Results from other searches
are also shown in the $\alpha-\lambda$ plane
in this figure.  The exclusion curve labeled as 
Irvine \cite{ho:85} is the result of an
experiment that used a torsion balance
to measure the force on a test mass suspended inside a hollow
cylinder; it provides the most stringent limits in the centimeter range.
The curve labeled by Moscow is obtained from a torsion balance
experiment which used a dynamic-resonant
technique to distinguish the gravitational interaction from
backgrounds \cite{mi:88,lo:99}.  The curve labeled by
Lamoreaux corresponds to
an  experiment that measured
the force between a flat, metal-plated quartz disk
and a metal-plated spherical lens \cite{lo:99,la:97}.
These Cavendish-type experiments are sensitive to
new forces weaker than gravity by a few orders of magnitude
at distances of a centimeter,  of  comparable strength as Newtonian 
gravity at distances of a few hundreds of
microns, and stronger than gravity by a few orders of magnitude
at distances of order tens of microns.  In addition, data from
experiments that measure the
Casimir force between uncharged conducting
surfaces \cite{ca:48} can also probe a  Yukawa interaction.
These Casimir-type experiments
are sensitive to interactions many orders of magnitude
stronger than gravity at the nanometer regime. 

The predictions and allowed regions in the $\alpha-\lambda$
parameter space
from other theoretical considerations are also 
presented in the figure; they include scenarios with axions, 
dilatons and scalar moduli fields from string theory, and
attempted solutions to the cosmological constant problem.

Lastly, the Randall-Sundrum non-compact extra dimension scenario
with small warping, $1/k\approx 10$\, mm, predicts  multiplicative
corrections of $(1+O(10^{-2}))$ to the Newtonian potential.  These
modifications are also detectable in short range
gravity experiments and are distinguishable from the predictions of
compact extra dimensions \cite{chu:00}.

%% file: astro.tex
Astrophysical and cosmological considerations impose strict
constraints on some theories of extra dimensions; in particular,
early universe cosmology can be drastically altered from the standard
picture.  The typical
energy scale associated with such considerations is of order 100 MeV,
and models with KK states that can be produced in this energy regime 
are highly restricted. 

For the case of large extra dimensions of
flat and toroidal form discussed in Section 2.1, the astrophysical
bounds far surpass those from collider or short range gravity
experiments for $\delta=2$.  If these large additional dimensions
are compactified on a hyperbolic manifold instead, then the 
astrophysical constraints are avoided \cite{Kaloper:2000jb} 
as the modified spectrum of KK graviton states admits for a 
first excitation mass of order several GeV.  Alternatively, these
bounds are also weakened if a Ricci term is present on the brane since
that serves to suppress graviton emission rates \cite{Dvali:2001gm}.
Similarly, bounds of
this type are not applicable to the Randall-Sundrum scenario
of warped extra dimensions with two branes since the first graviton
KK state occurs at a $\sim$TeV.  In the alternate
Randall-Sundrum model where the second brane is taken to infinity,
the resulting cosmological constraints are also found to be very mild
\cite{Hebecker:2001nv}.

We now describe the various astrophysical and cosmological
considerations that restrict the scenario with large flat extra
dimensions.  These processes include graviton emission
during the core collapse of supernovae, the heating of neutron
stars from graviton decays, considerations 
of the cosmic diffuse $\gamma$-ray background, overclosure of the
universe, matter dominated cooling of the universe, and reheating
of the universe.   The restrictions obtained from processes that
include effects from the decays of KK states rely on the assumption that
the KK modes can only decay into Standard Model particles on one
brane, {\it i.e.}, there are no additional branes in the theory,
and that decays into other KK modes with smaller bulk momenta
do not occur.

During the core collapse of type II supernova (SN),
most of the gravitational binding energy
is radiated by neutrinos. This hypothesis 
has been confirmed by measurements of neutrino
fluxes from SN1987A by the Kamiokande  and IMB 
collaborations \cite{kam:87}.
Any light, neutral, weakly interacting particle which couples
to nucleons, such as bulk gravitons, will compete with neutrinos in 
carrying energy away from the stellar interior. 
The rate at which the supernova core can lose energy through 
emission of KK states can then be used to constrain
the fundamental scale $M_{D}$ \cite{cu:99,hanh:00}.
The graviton emission process is nucleon `gravisstrahlung', 
$N + N \rightarrow N + N + X$, where $N$ can be a proton or neutron, 
and $X$ represents the contributions from massive KK graviton states,
ordinary gravitons, and the KK dilaton (scalar) modes which are a 
remnant of the bulk graviton decomposition.
If present, this gravisstrahlung process 
would provide an additional heat sink and accelerate 
the supernova cooling in violation with the observations of
SN1987A.  This process is highly dependent on
the temperature of the core at collapse, which is estimated to be 
$T\approx 30-70$ MeV, and on the core density, 
$\rho\approx(3-10)\times 10^{14} {\rm g cm}^{-3}$.  Several additional
uncertainties, such as the form of the nucleon scattering matrix,
the specific heat of matter at high density, and the neutrino transport
mean-free path at high density, also enter the calculation.
These uncertainties  are computed using the well studied
nucleon-nucleon axion bremsstrahlung process.  The most conservative
constraint on KK emission \cite{hanh:00}  yields  
$R_c\le 7.1\times10^{-4}$ mm for $\delta=2$ and $R_c\le 8.5\times10^{-7}$ mm
for $\delta=3$, taking $T_{SN1987A}=30$ MeV. 

A complementary bound arises from the radiative decay of the Kaluza Klein
gravitons produced by the core collapse of all supernovae that have
exploded during the history of the universe (SNe).
The two photon decay mode is kinematically 
favored for the lower mass KK modes \cite{Han:1999sg}
with this lifetime being $\tau_{\gamma\gamma}\approx  3\times10^{9} 
\rm yr \left( 100~MeV\over m_{KK} \right)^{3}$. Over the 
age of the universe, a significant fraction of the KK states emitted
from supernovae cores will have decayed into photons, contributing to 
the cosmic diffuse $\gamma$-ray background. This is estimated using the
present day supernova rate and the gravisstrahlung rate discussed above.
A bound on the size of the additional dimensions is then 
imposed from the measured cosmic $\gamma$-ray
background. For a choice of cosmological parameters 
the predicted $\gamma$-flux exceeds the observations 
by EGRET or COMPTEL \cite{gflux}  unless the fraction of the SN energy
released via gravisstrahlung is less than about 0.5-1\% of the total.
For two extra dimensions, the limit on the compactification radius
is $R_c\le 0.9 ~10^{-4}$ mm   
and for three extra dimensions the bound is $R_c\le 0.19~ 10^{-6}$ mm 
\cite{han:01}.  
Additional contributions to the cosmic diffuse $\gamma$-ray
background arise when the KK gravitons are produced from other sources
such as neutrino annihilation, $\nu\bar\nu\to G_n\to\gamma\gamma$.  These 
were considered in \cite{Hall:1999mk}, and by placing a bound on the
normalcy temperature required by Big Bang Nucleosynthesis the limit
$R_c\le 5.1\times 10^{-5}$ mm for $\delta=2$ is obtained.

In \cite{Hall:1999mk}, it is assumed that the universe enters the 
radiation dominated epoch instantaneously at the reheating
temperature. However, it is plausible that the universe enters the 
radiation epoch after being reheated by the decay of a massive scalar
field or by some other means of entropy production.
If a large number of KK states  are produced during reheating,
they are non-relativistic and hence are not diluted
by entropy production.  Their subsequent decays contribute
to the diffuse $\gamma$-ray background.  Using data
from COMPTEL and EGRET \cite{gflux}, 
the constraints on  $M_{D}$ are tightened and 
are  167, 21.7, 4.75, and 1.55 TeV for 
 $\delta=2,3,4,$ and 5, respectively, 
assuming that a 1 GeV maximum temperature
is reached during reheating \cite{han_2:01}. 

The escape velocity of a neutron star is similar to the average speed
of thermally produced KK states in a SN core collapse, and hence 
a large fraction of the KK states can become trapped within the core
halo.  The decays of these states will continue to be a source of  
$\gamma$-rays long after the SN explosion.  Comparisons of the
expected contributions to the $\gamma$-ray flux rate from this source
with EGRET data \cite{gflux} from nearby neutron stars and pulsars constrains
\cite{Hannestad:2001xi} the  fraction of the SN energy
released via gravisstrahlung to be less than about $10^{-5}$
of the total.  For two extra dimensions this yields the bound
$M_D\gsim 450$ TeV and for $\delta=3$ the constraint is $M_D\gsim 30$
TeV.  The expected sensitivity from GLAST \cite{bloom} 
will increase these limits
by a factor of 2 to 3.

The Hubble space telescope has observed that the surface temperature
of several older neutron stars is higher than that expected in 
standard cooling models.  A possible source for this excess heat
is the decays of the KK graviton states trapped in the halo surrounding 
the star.  The $\gamma$'s, electrons, and neutrinos from the KK decays
then hit the star and heat it.  For the estimated heating rate from this
mechanism not to exceed the observed luminosity, the fraction of the SN energy
released via gravisstrahlung must be $\lsim 5\times 10^{-8}$
of the total \cite{Hannestad:2001xi}, with the exact number being
uncertain by a factor of a few due to theoretical and experimental
uncertainties.  This is by far the most stringent
constraint yielding $M_D\gsim 1700\,, 60$ TeV for $\delta=2\,, 3$,
respectively.  Although the calculations for SN emissions have not been
performed for $\delta>4$, simple scaling suggests that this mechanism
results in $M_D\gsim 
4\,, 0.8$ TeV for $\delta=4\,, 5$, respectively.

Once produced, the massive KK gravitons are sufficiently long-lived
as to potentially overclose the universe.  Comparisons of KK graviton
production rates from photon, as well as neutrino, annihilation to the 
critical density of the universe results \cite{Hall:1999mk} in
$R_c<1.5h\times 10^{-5}$ m for 2 additional dimensions, where $h$ 
is the current Hubble parameter in units of 100 km/sMpc.
While this constraint is milder than those obtained above, it is
less dependent on assumptions regarding the existence of additional
branes.

Overproduction of Kaluza Klein modes in the early universe could
initiate an early epoch of matter radiation equality which would
lead to a too low value for the age of the universe.
For temperatures below $\sim 100$ MeV,
the cooling of the universe can  be accelerated by 
KK mode production and evaporation into the bulk, 
as opposed to the normal cosmological expansion.
Using the present temperature of the cosmic microwave background  
of 2.73 K ($=2.35\times 10^{-10}$ MeV) and
taking the minimum age of the universe to be 12.8 Gyrs ($=6.2\times
10^{39}$ MeV$^{-1}$), as determined by the mean observed age of
globular clusters, a maximum temperature can be imposed at
radiation-matter equality which cannot be exceeded by the
overproduction of KK modes at early times.
The resulting lower bounds are
$M_{D}$ are 86, 7.4, and 1.5 TeV for $\delta=2,3,$ and 4 respectively
\cite{fai:01}.  
Further considerations of the effects from overproduction of KK
states on the characteristic scale of the turn-over
of the matter power spectra at the epoch of matter radiation equality
show that the period of inflation must be extended down to very
low temperatures in order to be consistent with the latest data
from galaxy surveys \cite{fai:02}. 

We collect the constraints from these considerations in 
Table \ref{astrocons}, where we state
the restrictions in terms of bounds on the fundamental
scale $M_D$.  We note that
the relation of the above
constraints to $M_D$ is tricky as numerical conventions, 
as well as assumptions regarding 
the compactification scheme, explicitly enter some of the computations;
in particular, that of gravisstrahlung production during  supernova
collapse.  In addition, all of these bounds assume that all of the
additional dimensions are of the same size.  The constraints in the 
table are thus merely indicative and should not be taken as exact.

To conclude this section, we discuss the possible contribution of
graviton KK states to the production of high-energy
cosmic rays beyond the GZK cut-off of $10^{20}$ eV.  
About 20 super-GZK events have been
observed and their origin is presently unknown.
In the case of large extra dimensions, KK graviton exchange can
contribute to high-energy $\nu$-nucleon scattering and produce hadronic
sized cross sections above the GZK cut-off 
for $M_D$ in the range of 1 to 10 TeV \cite{Jain:2000pu}.
In the Randall-Sundrum model of localized gravity, neutrino
annihilation within a GZK distance of the earth can produce a 
single graviton KK state on resonance which subsequently decays
hadronically \cite{dhr_astro}.  For neutrinos of mass $m_\nu\sim
10^{-2}$ to $10^{-1}$ eV, and graviton resonances of order a TeV,
super-GZK events can be produced.
Under the assumption that the incident 
neutrino spectrum extends in neutrino energy with a reasonably 
slow fall-off, the existence of a series of 
$s$-channel KK graviton resonances will lead to a series of ultra-GZK
events.  The rates for these bursts are 
generally at or near the present level of observability for a wide range of 
model parameters. The fact that such events are not as yet observed can be 
used to constrain the parameter space of this model once a specific form of 
the neutrino energy spectrum is assumed. 

\begin{table}
\centering
\begin{tabular}{|l|c|c|c|c|} \hline\hline
 & \multicolumn{4}{c|} {$\delta$} \\ \hline
 & 2 & 3 & 4 & 5 \\ \hline
Supernova Cooling \cite{hanh:00} & 30 & 2.5 & & \\
Cosmic Diffuse $\gamma$-Rays: & & & & \\ 
\quad\quad Cosmic SNe \cite{han:01} & 80 & 7 & &\\
\quad\quad $\nu\bar\nu$ Annihilation \cite{Hall:1999mk} & 110 & 5 & &\\
\quad\quad Re-heating \cite{han_2:01} & 170 & 20 & 5 & 1.5 \\
\quad\quad Neutron Star Halo \cite{Hannestad:2001xi} & 450 & 30 & &\\
Overclosure of Universe \cite{Hall:1999mk} & $6.5/\sqrt h$ & & & \\
Matter Dominated Early Universe \cite{fai:01} & 85 & 7 & 1.5 & \\
Neutron Star Heat Excess \cite{Hannestad:2001xi} 
& 1700 & 60 & 4 & 1 \\ \hline\hline
\end{tabular}
\caption{Summary of constraints on the fundamental scale 
$M_D$ in TeV from astrophysical and
cosmological considerations as discussed in the text.}
\label{astrocons}
\end{table}

%% file: sum.tex
If the structure of spacetime is different than that readily 
observed, gravitational physics, particle physics and cosmology
are all immediately affected. The physics of extra dimensions 
offers new insights and solutions to fundamental questions 
arising in these fields. Novel ideas and frameworks are continuously born and
evolved. They make use of string theoretical features and tools and 
they may reveal if and how the
11-dimensional string theory is relevant to our four-dimensional
world.  One such recent idea is that
the extra dimension(s) are generated 
over a finite energy region while the theory is
4-dimensional both at high- and low-energies
\cite{deconstruct}. The framework for this scenario does not presently
include gravity, but work is underway to do so. 
In this theory, the extra dimension is generated
dynamically and  ``deconstructed'' at both high- and low-energies. 
A conceptually  similar idea is that of a
correspondence between 5-dimensional Anti-de-Sitter space and
a 4-dimensional conformal field theory 
(AdS/CFT correspondence) \cite{adscft}.

If the fundamental scale of gravity is at roughly a TeV, 
then future colliders will
directly probe new exotic degrees of freedom in addition to
the Kaluza Klein modes of extra dimensions, including the effects
of quantum gravity itself. We do not yet have
unambiguous predictions for this new and unknown physics, but it could
take the form of  new
strongly interacting gauge sectors or string or brane excitations. 
For example, the exchange of string
Regge excitations of Standard Model particles in $2\to 2$ scattering
\cite{Cullen:2000ef,dudas}  would appear as a contact-like
interaction, similar to that of graviton KK exchange, but with a
large strength.  

It is possible that inelastic scattering at energies $\gg$ TeV
 could be dominated by the production of strongly coupled objects
such as  microscopic black holes  \cite{BH}. Assuming that these decay 
via Hawking radiation, they
would then be observable in future very
high-energy colliders \cite{Rizzo:2002kb}.

In this review we have summarized the pioneering frameworks
with extra spatial dimensions which have observable consequences at
the TeV scale. We have outlined some of the experimental 
observations in particle and gravitational physics as
well as astrophysical and cosmological considerations 
that can constrain or confirm these
scenarios. These developing ideas and the wide interdisciplinary 
experimental program that is charted out to 
investigate them  mark a renewed effort to describe the 
dynamics behind spacetime.